\documentclass[useAMS]{mn2e}
\usepackage{epsfig}
\usepackage{amsmath}
\usepackage{amssymb}
\usepackage{graphics}

\title[The formation of He-rich sdO stars]{On the formation of single
  and binary helium-rich sdO stars} \author[Justham, Podsiadlowski {\rm\&}
  Han]{Stephen Justham,$^{1,2}$\thanks{Email:
    sjustham@pku.edu.cn}, Philipp Podsiadlowski$^{2}$ \& Zhanwen
  Han$^{3,4}$ \\
 $^{1}$ The Kavli Institute for Astronomy {\rm\&}
Astrophysics, Peking University, Beijing 100871, China \\ 
$^{2}$ Sub-department of Astrophysics, Oxford University, Keble Road,
  Oxford, OX1 3RH, United Kingdom \\ 
$^{3}$ National Astronomical Observatories / Yunnan Observatory, 
Chinese Academy of Sciences, Kunming 650011, China \\
$^{4}$ Key Laboratory for the Structure and Evolution of Celestial Objects, Chinese Academy of Sciences}

\date{Submitted 1 March 2010; Accepted 6 August 2010}

\pagerange{\pageref{firstpage}--\pageref{lastpage}}
\pubyear{2010}

\begin{document}
\maketitle

\label{firstpage}

\begin{abstract}
We propose a formation channel for the previously unexplained
helium-rich subdwarf O (He-rich sdO) stars in which post-subdwarf B (sdB)
stars (i.e.\ hybrid COHe white dwarfs) reignite helium
burning in a shell after gaining matter from their helium white-dwarf (WD)
companions.  Such short-period binaries containing post-sdB WDs and
helium WDs are predicted by one of the major binary formation
channels for sdB stars.  In the majority of cases, mass transfer is
expected to lead to a dynamically unstable merger event, leaving a
single-star remnant.  Calculations of the evolution of
these stars show that their properties are consistent with the
observed He-rich sdO stars.  The luminosity of these stars is
about an order of magnitude higher than that of canonical sdB stars.
We also suggest that binary systems such as PG 1544+488 (Ahmad et
al.\ 2004) and HE 0301-3039 (Lisker et al.\ 2004), which each contain
two hot subdwarfs, could be the outcome of a double-core
common-envelope phase. Since this favours intermediate-mass progenitors,
this may also explain why the subdwarfs in these systems are
He-rich. 
\end{abstract}


\section[]{Introduction}

In recent years there has been an extensive debate over the origin of hot
subdwarf stars, largely because they seem to provide the best
explanation for the UV-upturn seen in elliptical galaxies (e.g.\ Yi,
Demarque \& Oemler 1997, 1998; Han, Podsiadlowski \& Lynas-Gray
2007). The production of such stars in globular clusters, where they
are often referred to as \emph{extreme horizontal branch} (EHB)
stars, has been a long-standing mystery (see, e.g., van den
Bergh 1967; Sandage \& Wildey 1967; Soker 1998; Han 2008).

Subdwarf B (sdB) stars are believed to be helium-core-burning stars
with masses $\rm \sim 0.5~M_{\odot}$, posessing very small ($\rm <
0.02~M_{\odot}$) hydrogen-rich envelopes (Heber 1986; Saffer et
al.\ 1994). Their formation has been studied in some detail by Han et
al.\ (2002 \& 2003), who showed that binary evolution can account for
the properties of the observed sdB stars. Other authors have suggested
enhanced mass-loss from single red-giant stars in order to try to
explain the formation of sdB stars (e.g.\ Yi et al.\ 1997; see also
Han, Podsiadlowski \& Eggleton 1994), or mixing driven by a very late
helium flash (Sweigart 1997; Brown et al.\ 2001), or that interactions with
planets can eject the envelopes of red giants (Soker 1998;
Nelemans \& Tauris 1998).

Subdwarf O (sdO) stars are assumed to be related to sdB
stars. Stroeer et al.\ (2007) have examined the formation of sdO stars
and concluded that the sdO stars with a subsolar photospheric helium abundance
(`helium-deficient') have a different origin to the `helium-enriched' (He-rich)
sdO stars (which they define as having a super-solar helium
abundance). Specifically, Stroeer et al.\ found that the
helium-deficient sdO stars are likely to be evolved sdB stars, but
that the He-rich sdO stars cannot be explained through the
canonical evolution of sdB stars.\footnote{Zhang, Chen \& Han (2009) have also
 tried to constrain the fraction of sdO stars which can simply be evolved sdB stars 
and which fraction requires another explanation.} Proposed formation
channels for those He-rich sdO stars include mergers of two helium white dwarfs
(Saio \& Jeffery 2000) and the `hot flasher scenario' (e.g.\ Moehler
et al.\ 2007; Miller Bertolami et al.\ 2008).  Though neither of these
models seems entirely satisfactory, one piece of evidence which
apparently favours the white-dwarf merger scenario is the very low
binary fraction of the He-rich sdO stars (see, e.g., Heber et
al. 2006; Heber 2008). Here we show that the existence of
He-rich sdO stars, including single ones, is a natural
consequence of one of the binary formation channels for sdB
stars.\footnote{Our proposed formation channel seems unable to produce
  the class of He-rich sdB stars (Naslim et al.\ (2010); we discuss
  the implications of this in section \ref{sec:HeWDmergers}.}

Section \ref{sec:Explanation} explains the proposed formation channel,
where section \ref{sec:Evolutions} presents evolutionary
calculations and section \ref{sec:Populations} compares population
synthesis expectations to the observed sample of He-rich sdO
stars. Section \ref{sec:DoubleCore} then introduces a new formation
channel for double hot subdwarfs and examines its possible
implications.

\section[]{A new merger channel} 
\label{sec:Explanation}

The merger of a helium white dwarf (WD) with a post-sdB star is
predicted in one of the important binary channels for the formation
of sdB stars (Han et al.\ 2002; 2003), but has not been considered
in detail before. Here we propose that such mergers are the natural
progenitors of He-rich sdO stars.

\begin{center}
\begin{figure}
\begin{center}
\epsfig{file=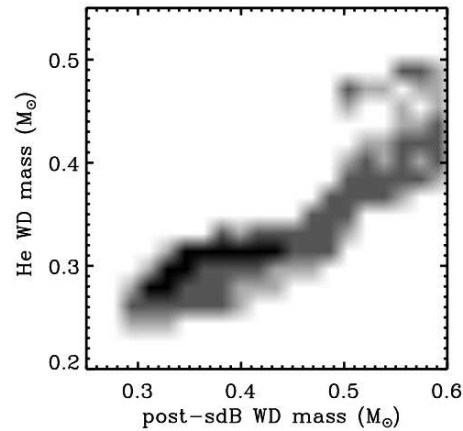, width=6cm}
\end{center}
\caption{
\label{fig:initdist}
The pre-merger distribution of the component masses of binary
  systems expected to produce He-rich sdO stars (taken from the
  preferred binary population synthesis model of Han et al. [2003]).
  Darker colours indicate more systems. Once the mass of the sdB star
  approaches 0.6 $\rm M_{\odot}$, the outcome of the mass-transfer
  episode is better described by the CO+He WD merger calculations of
  Saio \& Jeffery (2002). The stellar evolution calculations in this
  paper use representative post-sdB masses of 0.35, 0.4 and 0.46 $\rm
  M_{\odot}$.  }
\end{figure}
\end{center}

Han et al.\ (2002 \& 2003) showed that a significant fraction
of sdB stars is expected to exist in short-period binaries
containing a helium WD (see Fig.~\ref{fig:initdist}). This is the result of the `second
common-envelope channel', where a WD spirals into the giant envelope
of the sdB progenitor, thereby ejecting that envelope.  After the
helium-burning phase of the sdB star has been completed, the remnant
will become a hybrid WD with a carbon-oxygen (CO) core and a thick
helium (He) envelope. If the
orbital period is short enough,\footnote{Typically $\lesssim$ 7 hr for interaction 
within a Hubble time; see footnote \ref{fn:Pmerge}.}
gravitational radiation will drive the binary components towards
each other until eventually mass transfer is initiated from the
lighter helium WD onto the post-sdB He-CO WD star. Once a sufficiently thick
helium shell has been built up around the CO core, helium will
(re-)ignite in this shell. In the following section, we will show
that such objects have all the main observational characteristics of
He-rich sdO stars.

Following the pioneering work of Saio \& Nomoto (1998),
Saio \& Jeffery (2000, 2002) simulated the merger of white dwarfs
(first of two He WDs and later of a 0.6 $\rm M_{\odot}$ CO WD
accreting from a He WD).  The situation we consider here is very
similar; even though our calculations in section \ref{sec:Evolutions}
do not treat the mass-transfer/merger phase, the work of Saio \&
Jeffery strongly suggests that the post-merger star will undergo
stable helium burning, with weak He shell flashes during the accretion
phase. We cannot completely exclude the possibility of catastrophic
explosive He burning that could destroy the star or eject a large
fraction of the accreted envelope, but the accretion rates in this
case are significantly higher than those predicted to produce `.Ia'
supernovae (see, e.g., Bildsten et al.\ 2007; Shen \& Bildsten 2009). 
However, since we cannot properly calculate the burning during the
accretion phase, we cannot show how the newly-reignited systems
behave during their approach to equilibrium.\footnote{Note that the details of the ignition phase should
depend on the accretion rate (see, e.g., Saio \& Nomoto 1998).}  In principle, some of the
observed He-sdOs could be systems which have not yet attained their
thermal equilibrium structure.

Not all of these systems should experience dynamically unstable mass
transfer.  The results of Han \& Webbink (1999) indicate that the mass
transfer from the helium white dwarf in these systems will usually be
dynamically unstable, leading to a merger and ultimately a single
He-rich sdO star. This may naturally explain the high fraction of
apparently single He-rich sdO stars (see, e.g., Stroeer et
al.\ 2007).  However, the systems which undergo a stable mass-transfer
phase could potentially leave binary He-rich sdO stars. 
In this case, the sdO star could still be in a close binary with an
orbital period $\lesssim 2\,$hr and be accreting at a low rate from
a potentially very low-mass companion (i.e.\ a few 0.01\,M$_{\odot}$; 
see, e.g., Fig.~7 in Rappaport, Podsiadlowski \& Horev [2009]; also see
the case of the millisecond pulsar PSR 1957+20 [Fruchter et al.\ 1988]).

\begin{figure*}
\epsfig{file=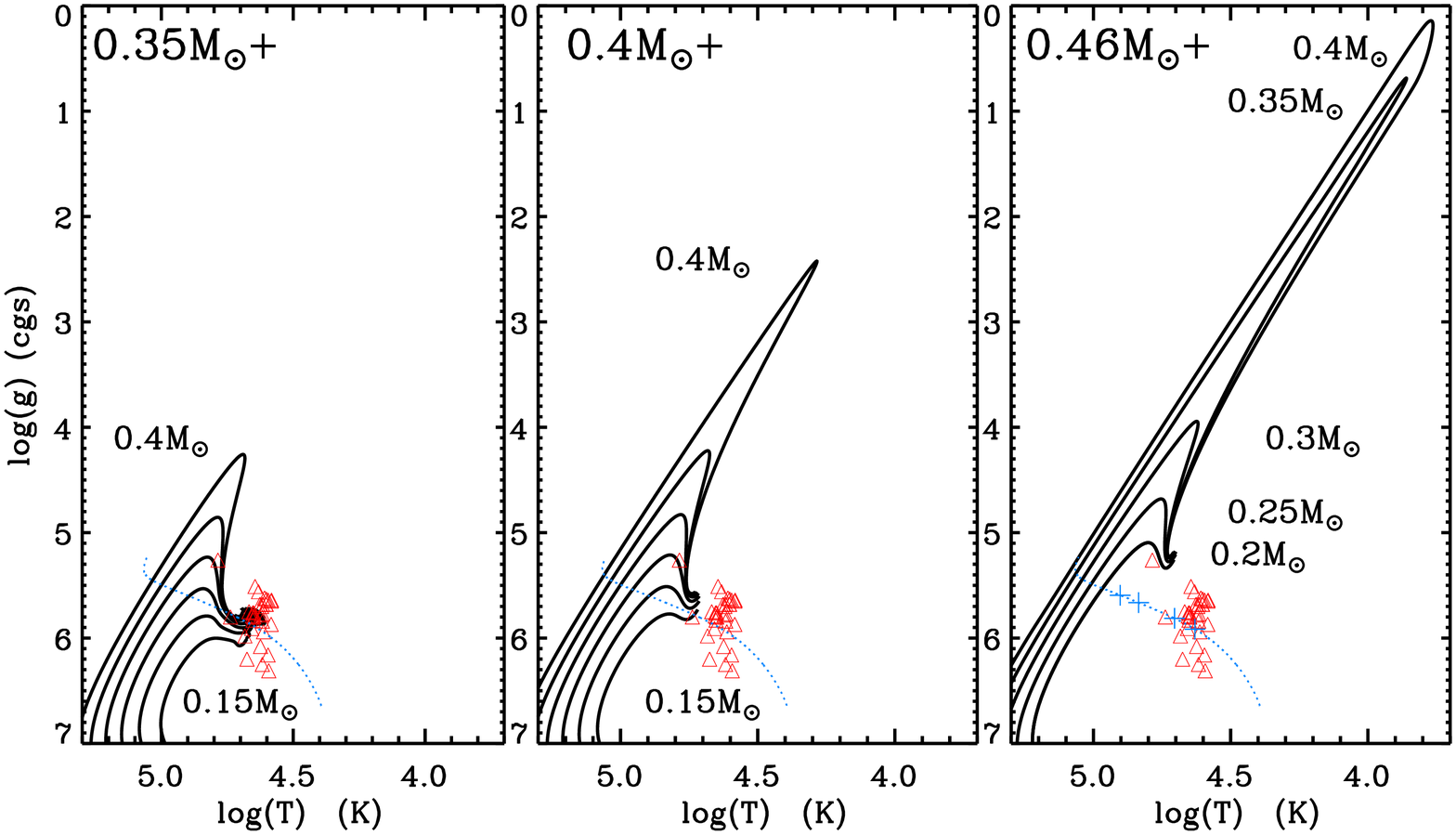, width=16cm}
\epsfig{file=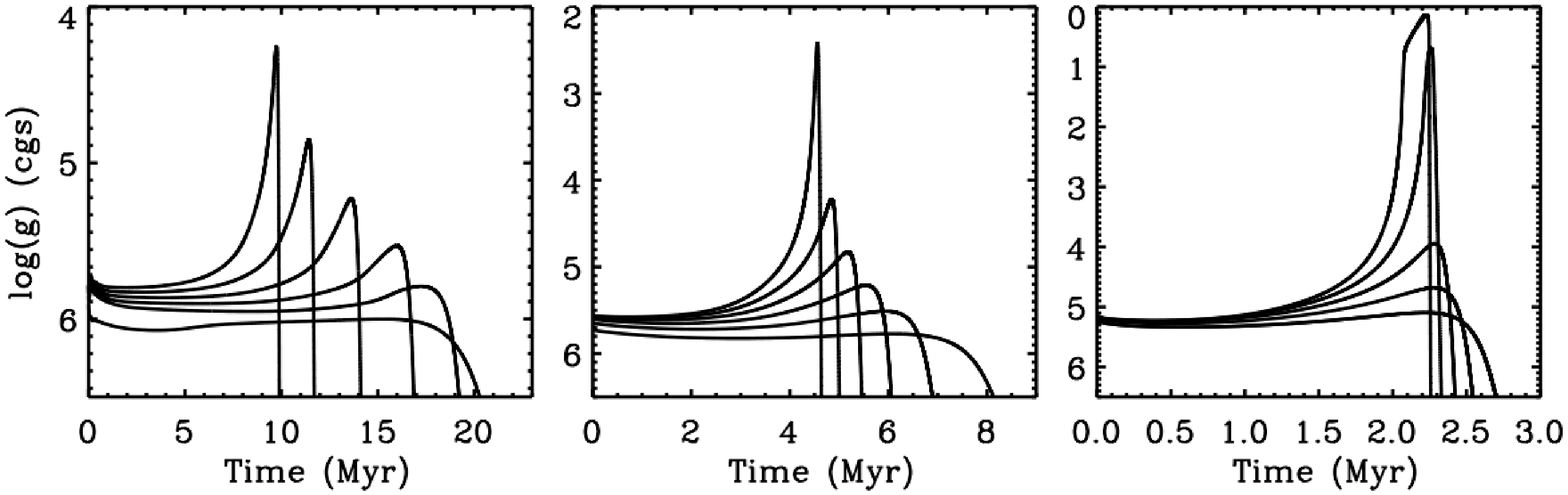, width=16cm}
\caption{
\label{fig:logg_tracks}
The post-merger evolutionary tracks of stars formed by mergers of
post-sdB WDs with helium WDs. The panels show the tracks for mergers
of a 0.35 M$_{\odot}$ (left), 0.4 M$_{\odot}$ (centre) and 0.46
M$_{\odot}$ (right) post-sdB star, respectively, with a range of
different helium masses (in steps of 0.05 M$_{\odot}$, as
labelled). The red triangles show the locations of the He-sdOs given
in Stroeer et al.\ (2007). The dotted curve in each upper panel indicates
our theoretical helium main sequence; in the right hand panel we mark
helium main sequence masses of 3, 2, 1 and $0.7~\rm M_{\odot}$
(left--right) with plus symbols. The lower row of panels shows the time
evolution of the surface gravity of these stars. Note that they spend
the majority of their lifetimes near to the start of the tracks, close
to the observed He-sdO stars, especially for the stars formed from the
0.35 M$_{\odot}$ post-sdB stars.}
\end{figure*}

\begin{figure*}
\epsfig{file=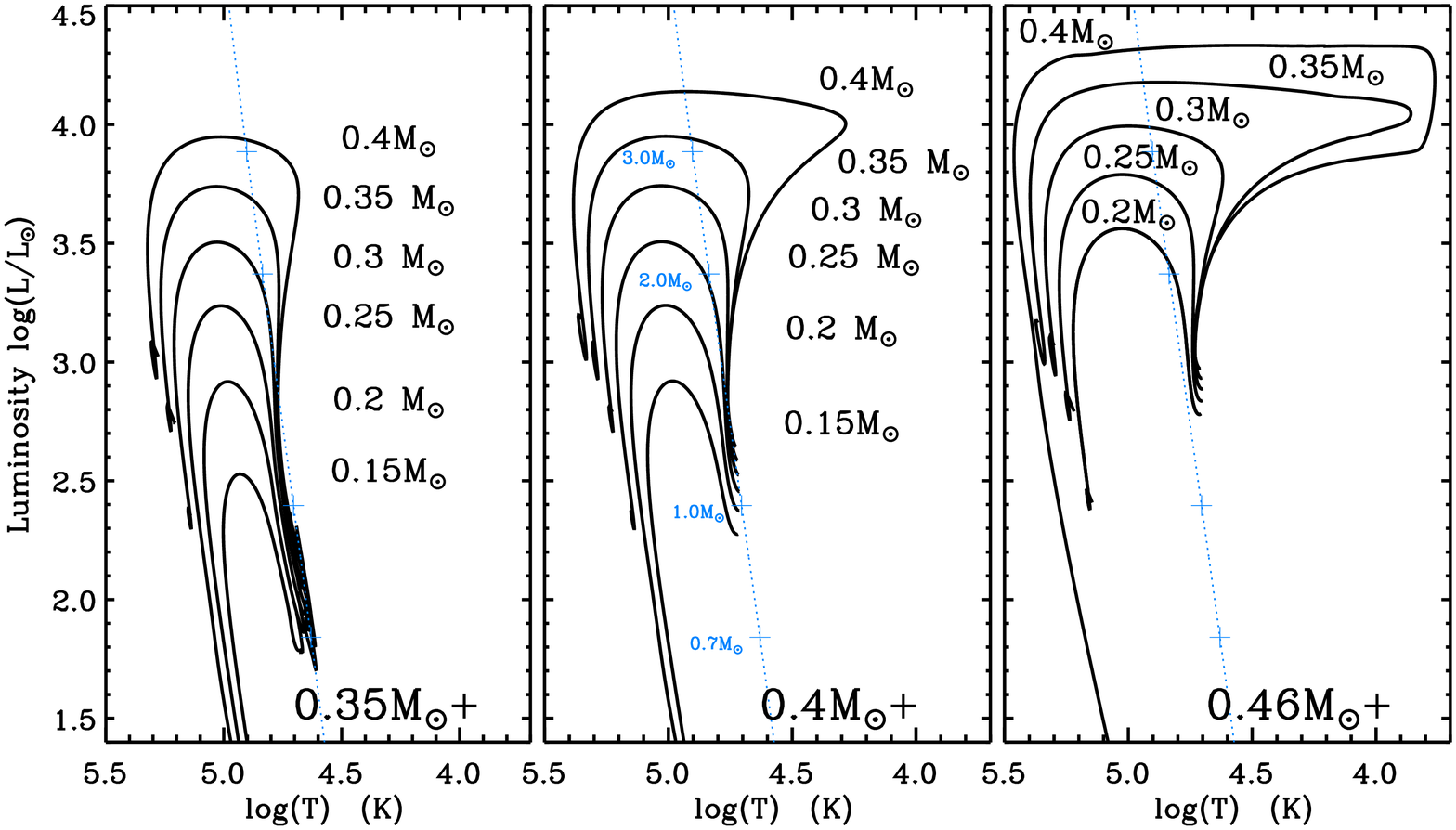, width=16cm}
\epsfig{file=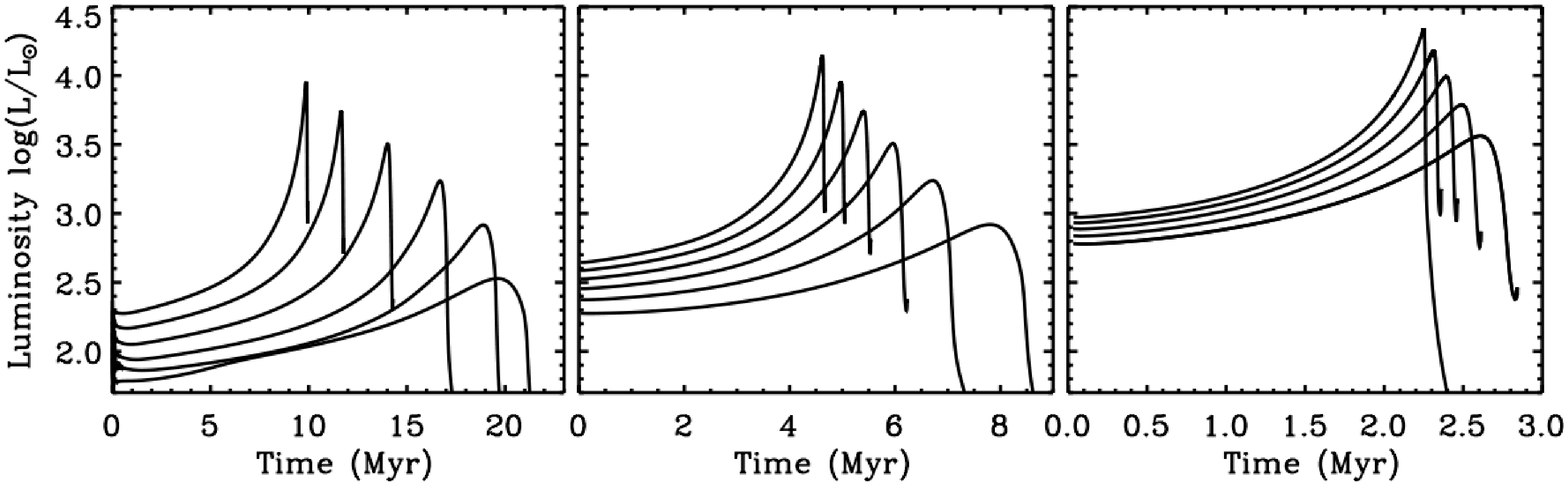, width=16cm}
\caption{
\label{fig:HR_and_times}
The same stellar calculations as in Fig.~\ref{fig:logg_tracks}, but
this time showing the luminosity evolution of the stars. The overall
evolution in the HR diagrams is from right to left. These merger
products are significantly more luminous than canonical sdB
stars. The dotted curve in each upper panel shows our theoretical 
helium main sequence; helium main sequence masses of 
3, 2, 1 and $0.7~\rm M_{\odot}$ are marked with plus symbols.}
\end{figure*}

\subsection[]{Post-merger calculations}
\label{sec:Evolutions}

For our stellar evolution calculations we used Eggleton's stellar
evolution code (Eggleton 1971; Pols et al.\ 1995) with a metallicity
of 0.02 along with the convective overshooting calibration of Pols et
al.\ (1998). We first evolved a set of sdB stars using sdB models from
Han et al.\ (2002, 2003) in order to produce a set of post-sdB models
onto which matter was added from an assumed helium WD companion.

For the sdB stars which undergo shell burning, we generally found it
necessary to perform the merger very late in the growth of the sdB
core, but before the helium burning had completely finished. This
avoids having to ignite helium degenerately, though we still regularly
found steep and numerically troublesome increases in the nuclear
energy generation rate as we added the matter from the He WD. During
the merger process, we artificially switched off the nuclear evolution
of the object (the energy generation continued as normal but the composition was frozen). 
To some extent, this enabled us to add the mass as slowly as
was required for our models to converge; we typically added mass at a
rate that was over an order of magnitude smaller than the
Eddington-limited accretion rate. Although the initial thermal
structure of these objects will not be a perfect match to that of the
actual merger products, it will adjust itself on a thermal
timescale. Hence we do not expect the subsequent nuclear evolution of
the stars to be significantly affected.

Our 0.35 $\rm M_{\odot}$ sdB stars are so cold and degenerate by the
time that they have completely exhausted their core He that we could
not follow mergers when starting from those structures. 
For these 0.35 $\rm M_{\odot}$ objects, we were able to take less degenerate 
models ($\psi_{\rm core} \approx 4$, where $\psi$ is the degeneracy parameter
[see, e.g., Kippenhahn \& Weigert 1990]) with almost no core helium
(${\rm log}~Y_{\rm core} < -8$) and set the core helium fraction to
zero by hand. We could then add matter to the surface of these objects
and follow the helium burning and subsequent evolution. When we tried to
perform mergers using stellar structures where the core helium had not
been reduced to zero, convection currents were driven by the residual
core helium burning, leading to significant mixing throughout the
star.

We investigated whether the age of the post sdB WD affected its
post-merger evolution, i.e. whether the core temperature and
degeneracy parameter $\psi$ at the time of the merger were
significant. This is not easy to study faithfully as our calculations
often fail to converge when the point of helium ignition has become
even mildly degenerate ($\psi \gtrsim2$). However, we adopted an
artifical way to test this. For a merger product which was not
degenerate enough to cause numerical problems, we switched off the
nuclear evolution whilst allowing the helium-burning energy generation
to continue. We then forcibly cooled the core as much as possible. For
lower helium WD masses (e.g. 0.2 and 0.25 $\rm M_{\odot}$), we were
able to reach core temperatures as low as $10^{6}$ K, and in all cases
below $10^{7}$ K. The initial core temperature appeared to have no
discernible effect on the long-term evolution of the merger product
once the artificial cooling was removed; the core temperature
increased again due to heat flowing inwards from the burning shell. We
also note that compressional heating of the accreting star during the
merger should reduce the degeneracy of the merger products compared to
our hydrostatic calculations, potentially further reducing any small
effect of the core temperature at the time of the merger. However, 
we cannot guarantee that this effect can be totally ignored; similar 
calculations have been performed by Iben (1990), who argued 
that the prior thermal history of the core was not negligible in the 
evolution of the post-merger star.

The results of our calculations are shown in Figs.
\ref{fig:logg_tracks} and \ref{fig:HR_and_times}. The majority of the
helium-burning phase of the merger products is spent close to the
majority of known He-sdO stars in the effective temperature -- surface
gravity diagram. This is particularly true for mergers of the
0.35$\rm~M_{\odot}$ post-sdB stars. Furthermore, the trend from the
0.4 to 0.35$\rm~M_{\odot}$ models suggests that the lowest mass sdB stars in
the population ($\approx 0.3\rm~M_{\odot}$; see Fig.
\ref{fig:initdist}) might help to fill in the low-temperature end of
the distribution. Pronounced helium-giant phases are experienced late
in the evolution of the merger products when the total mass exceeds
$\approx 0.8$\rm~M$_{\odot}$, in agreement with the behaviour described
by Trimble \& Paczynski (1974) for low-mass He stars.\footnote{Note that
these helium giants may experience a large amount of wind mass loss not
included in our calculations.}

\begin{figure*}
\epsfig{file=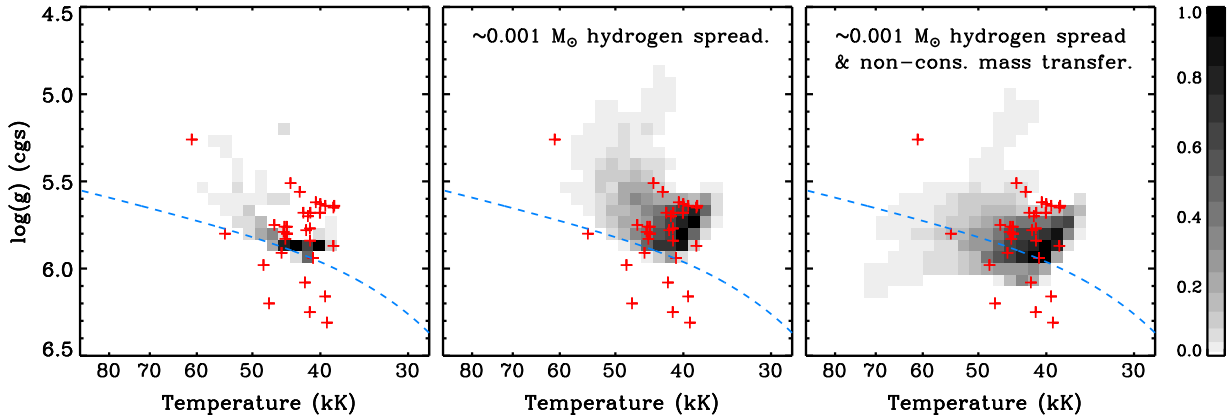, width=16.5cm}
\caption{
\label{fig:BPS}
  Population synthesis results for single He-rich sdOs, showing
  the distributions of surface gravity $g$ and effective temperature;
  darker regions indicate a higher density of systems. The crosses
  represent the observed He-rich sdO stars listed by Stroeer et
  al.\ (2007). The broken curve in each panel shows our 
  theoretical helium main sequence.
  The left panel shows the outcome for our hydrogen-free
  grid of stellar tracks, weighted by the results for close binary
  sdBs of Han et al.'s (2003), assuming that the binary components
  merge completely. In the central panel, the appearance of that same
  population is modified to take into account trace amounts of
  hydrogen remaining in the stellar atmospheres ($\lesssim 0.001 \rm
  M_{\odot}$; see text for details). The right panel is similar to the
  central panel, but also assumes that the merger process is
  non-conservative and that 0.1~M$_{\odot}$ of the helium WD is not
  accreted. Only the small fraction of stars with the highest surface
  gravities seem difficult to explain, but those stars seem likely to 
  have their surface gravitites adjusted as atmospheric models improve 
  (see, e.g., Hirsch \& Heber [2009]).  }
\end{figure*}

\begin{figure}
\begin{centering}
\label{fig:Lhist}
\epsfig{file=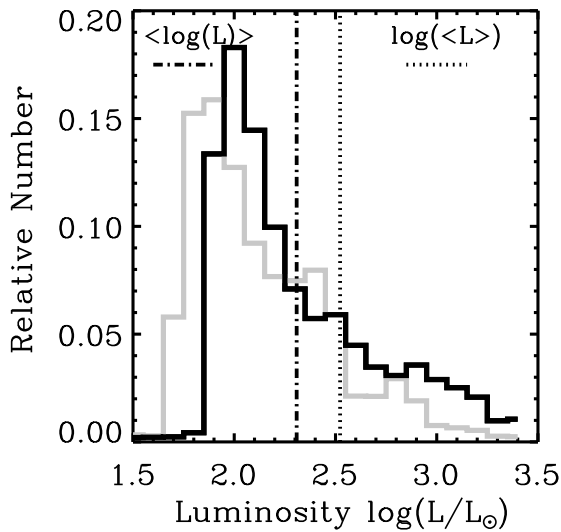, width=8cm}
\caption{ The black histogram shows the luminosity distribution taken
  from the population synthesis simulation which assumes that all of
  the He WD is accreted onto the newly-formed He-sdO star. Both the
  mean luminosities marked by vertical lines refer to that model. The
  light grey curve shows the distribution if we assume that 0.1 $\rm
  M_{\odot}$ of the He WD is somehow not accreted. In this case, the
  mean luminosity drops from $\rm \approx 300$~L$_{\odot}$ to $\rm
  \approx 200$~L$_{\odot}$.}
\end{centering}
\end{figure}

\subsection[]{Population synthesis}
\label{sec:Populations}

We have generated a library of evolutionary sequences for post-sdB + He
WD merger products, as described in the previous section. We then
combined that library with the sdB population predictions by Han et
al.\ (2003) to simulate the characteristic properties of the
  expected population of single He-rich sdO stars from this merger
channel.  Figure \ref{fig:BPS} compares the results of such a
population synthesis simulation to the properties of observed
He-rich sdO stars from Stroeer et al.\ (2007). We show the results
for two assumptions: one where all of the He WD is accreted by the
post-sdB star and one  where 0.1 $\rm M_{\odot}$ of the He WD is
lost from the system, to take into account possible systemic mass loss
or the formation of a disc around the merger product; a real
population is likely to have a more complicated combination of
conservative and non-conservative mass-transfer histories. The
correspondence between our models and the observations is quite good,
especially when we apply a correction to mimic the effects of small
amounts of hydrogen in the atmospheres of the sdO stars.

Our initial sdB models were completely hydrogen-free.
However, even a thin hydrogen envelope layer significantly
changes the appearance of a hot subdwarf (Han et al.\ 2002), making
it cooler and reducing its surface gravity. To take this into
account, Fig.~\ref{fig:BPS} shows the population synthesis
results both with and without an ad-hoc correction for trace
amounts of hydrogen, as estimated from figure 2 of Han et
al.\ (2002). We chose to uniformly spread each system in the $
T_{\rm eff}$--$\log g$ plane over a range given by the vector $-5000$
K in $T_{\rm eff}$ and 0.3 in $\log g$.\footnote{This is consistent
  with figure 2 of Han et al.\ (2002) for a H-rich envelope mass of
  0.001 $\rm M_{\odot}$, i.e. we assumed a uniform spread in H
  content from zero to $\approx 0.001 \rm M_{\odot}$. In reality,
  both the magnitude and direction of this vector should change
  slightly with effective temperature.}

We note that Stroeer et al.\ (2007) find that He-rich sdO stars
show higher carbon and nitrogen abundances than other sdO stars. This
seems consistent with our proposed formation channel, as mixing during
the merger phase could mix out material from the core of the post-sdB
star. However, the effect of these surface abundance changes 
has not completely been taken into account in our stellar
evolution calculations. These abundances may also affect the
atmospheric modelling of these stars (e.g. Lanz et al.\ 2004 \&
Stephan Geier, priv.\ comm.), hence some of the parameters of the
observed He-sdO stars may change as the atmospheric models
improve.\footnote{For example, Lanz et al.\ (2004) state that
  composition effects can produce a systematic error of 0.5 dex in the
  derived surface gravity of PG 1544+488.}  It seems likely that our
model could, in principle, account for most of the He-sdOs above the
helium main sequence. However, several He-sdOs below the He main
sequence can apparently only be accommodated by our simulations if
their surface gravities decreased as a result of improved atmosphere
calculations; the work of Hirsch \& Heber (2009) suggests that such an 
adjustment is very likely to occur.

\subsection{Population estimates}

We now estimate the birthrate of single He-rich sdOs, using the binary
population synthesis calculations of Han et al.~(2003). The formation
channel which produces sdB stars in close binaries with helium WDs is
their `second common-envelope channel' (CE2).  Their two
common-envelope channels produce the sdB stars known to be in close
binaries, which constitute about half of the known systems (we will
denote the fraction of sdB stars which are in close binaries as
$f_{\rm bin}$). For their model population which best fits the Galaxy,
the CE2 birthrate is $2.98\times 10^{-3} {\rm yr}^{-1}$ and the CE1
birthrate is $8.62\times 10^{-3} {\rm yr}^{-1}$, i.e. $\approx 26$\%
of the close binary sdBs come from the CE2 channel.  However, for the
common-envelope parameters used in their preferred model, only
$\approx 10$\% of the binaries produced via the CE2 channel are close
enough to merge within 10 Gyr.\footnote{\label{fn:Pmerge}
  This requires a
    post-common-envelope period $P_{0} \lesssim 7$ hr, assuming that
    gravitational wave radiation is the only angular-momentum loss
    mechanism and taking representative component masses of 0.35 \&
    0.45 $\rm M_{\odot}$.}  Hence we estimate the birthrate in the
Galaxy of He-rich sdOs ($B_{\rm He-sdO}$) to be $\sim 3$\% of the
birthrate of close binary sdBs ($f_{\rm bin} B_{\rm sdB} $).

Given the birthrates, we can estimate the relative number of sdB and
He-sdO stars using their respective lifetimes. For the He-sdO
population, we find a mean lifetime ($\tau_{\rm He-sdO}$) of $\approx$
10 Myr and, for sdB stars, we adopt the canonical lifetime $\tau_{\rm
  sdB}\approx 100$ Myr. The number of stars in the He-rich sdO
population, $N_{\rm He-sdO}$, is then related to the number of sdB
stars, $N_{\rm sdB}$, according to
\begin{equation}
N_{\rm He-sdO} = f_{\rm bin} N_{\rm sdB} \frac{\tau_{\rm He-sdO}}{\tau_{\rm sdB}} \frac{B_{\rm He-sdO}}{f_{\rm bin} B_{\rm sdB}},
\end{equation}
i.e., using the estimates above,
\begin{equation}
N_{\rm He-sdO} \sim 0.5 \times N_{\rm sdB} \times 0.1 \times 0.03  \approx \frac{N_{\rm sdB}}{667}.
\end{equation}
Thus, there should be $\sim 700$ times as many sdB stars as He-rich
sdO stars in the \emph{intrinsic} population. However, since
  He-rich sdOs are on average much brighter, this need to be corrected
  for the different potential detection volumes in order to be able to
  derive an estimate for the relative number in the \emph{observed}
sample.  Figure \ref{fig:Lhist} shows the luminosity distribution of
the predicted He-sdO population.  While a representative luminosity
for a sdB star is $\sim 15 $~L$_{\odot}$ (Han et al.~2003), the mean
luminosity of the He-sdO population is $\sim 300 $~$L_{\odot}$ (see
figure \ref{fig:Lhist}). This gives a relative detection volume of
$\sim 20^{1.5} \approx 90$.\footnote{This assumes a homogeneous
  spherical distribution of objects, which is unlikely to be an ideal
  approximation for stars of this luminosity within the Galactic
  disc. In the other extreme limit of a cylindrical thin-disc
  population, the ratio of detection volumes would be 20.}  This
estimate implies that there should be one He-sdO star known for every
$\sim 700/90 \approx 8$ sdB stars.

This is clearly a very approximate number. It assumes that we know both the 
number of systems produced by the CE2 formation channel and the fraction of those 
systems which have short enough periods to merge within a Hubble time. Uncertainties
in the CE2 channel include the treatment of common-envelope evolution itself, the extent 
to which the first mass transfer phase is non-conservative and the criterion for dynamically 
unstable mass transfer.

However, our estimate for the relative numbers of He-rich sdOs and sdB
stars seems to be a reasonable match to the observations.
 We are not aware of samples that can be compared precisely, but Stroeer et al.\ (2007) list
33 He-rich sdO stars; in a companion paper,
Lisker et al.\ (2005) state that over 200 sdB stars have been
`analysed for atmospheric parameters'. Heber (2008) increases that
number to `several hundred' sdB stars. The
Palomar-Green survey (Green, Schmidt \& Liebert 1986) found
$\sim$ 1 He-sdO for every 4 sdBs.\footnote{In the SPY sample of
hot subdwarfs (Napiwotzki et al.\ 2004; Lisker et al.\ 2005; Stroeer
  et al.\ 2007), sdO stars are likely to be over-represented with
  respect to sdB stars due to the selection criteria for the target
  list (Stroeer et al.\ 2007); thus using the ratio of 33 He-sdOs to 76
  sdBs published by that survey would be misleading. We do not know to
  what extent other selection effects affect the relative numbers in
  the published sdB and He-sdO populations.}

\subsection{A comparison with He-WD mergers}
\label{sec:HeWDmergers}

There seems to be little doubt that He-WD+He-WD mergers (Saio \&
Jeffery 2000) and He-WD+post-sdB mergers (this work) both happen. The
potential difference between the outcomes of those events is worth
examining.  Han et al.\  (2002, 2003) and Han (2008) have argued that
double He-WD mergers can explain single \emph{H-rich} sdB stars,
whilst here we argue that He-WD+post-sdB mergers can produce
\emph{H-poor} He-sdO stars. 

The difference between the atmospheres of sdB and sdO stars should be
able to account for this. Groth, Kudritzki \& Heber (1985) examined
the occurence of convection zones in hot subdwarf stars as a function
of temperature and composition; they argued that the 
atmospheres of helium-rich sdO stars should be convective, whilst those of
helium-poor sdOs and sdB stars are mostly radiative (see also Heber
[2009]). Hence gravitational settling can operate in most sdB stars --
producing He-poor photospheres -- but not in He-sdOs. The effective
temperatures of the merger products thereby determine which remain
He-rich; the cooler merger products experience settling. The majority
of He-WD mergers simulated in Han et al.\ (2002) are significantly
cooler than the merger products in Fig.~\ref{fig:BPS}, consistent with
the He-WD mergers producing H-rich sdBs and the merger channnel
proposed in this work producing He-sdOs. 

However, if that explanation is correct then the hotter He-WD mergers
may also contribute to the He-sdO population. In addition, note that
the boundary between the atmospheric regimes found by Groth et al.\ is
more complex than simply a division between sdO and sdB stars; in
particular, their calculations found that the temperature boundary 
which allows a convective atmosphere becomes cooler at lower surface
gravities.\footnote{Compare, e.g., figure 7 of Groth et al.\ with the continuous extension of the He-sdO
population into the He-sdB stars in, e.g., figures 4 and 5 of Naslim et al.\ (2010).}
Hence we should also expect some He-sdB stars to be produced by He-rich
merger products which have convective atmospheres but are cool enough to be sdB
stars.\footnote{Perhaps some He-rich subdwarfs could also be observed whilst their
  atmospheres are still experiencing gravitational settling.} This
would be helpful to our model, since it seems
that our merger products are too hot to explain He-sdB stars. Naslim
et al.\ (2010) have stressed that it is logical to consider
the evolutionary status of He-sdB and He-sdO stars together; our
merger scenario seems to require a second population to explain the He-sdBs:
potentially some He-WD mergers account for the He-sdBs as well as for the
some of the He-sdOs.

One outstanding and, as yet, unexplained piece of evidence is the
observation that He-sdOs come in both nitrogen-rich and carbon-rich
classes, with a further subset enhanced in both C and N, whilst
He-poor sdOs are not C or N rich (Stroeer et al.\ 2007). If some He-WD
mergers produce He-sdOs (perhaps, e.g., the more massive He-WD
mergers) then they might conceivably produce one composition subclass
whilst the He-WD+post-sdB mergers produce another subclass. Saio \&
Jeffery (2000) argue that the outcome of their double He-WD merger
model could have CNO-processed material at the surface (i.e. N-rich),
whilst Saio \& Jeffery (2002) produce a C-rich star from their
He-WD+CO-WD merger (\emph{not} a He-sdO).\footnote{
    He-sdBs are generally N-rich, consistent with being He-WD mergers, but the
    few C-rich examples are more puzzling (see, e.g., Naslim et al.\ 2010).}
Our merger model is somewhere between those examples; 
it is not clear which range of surface abundances our merger scenario 
might produce.   The phase of accretion and ignition
seems likely to be important for imprinting the surface carbon and 
nitrogen abundances of He-sdOs; we encourage future work to
investigate this detail.

\begin{figure}
\epsfig{file=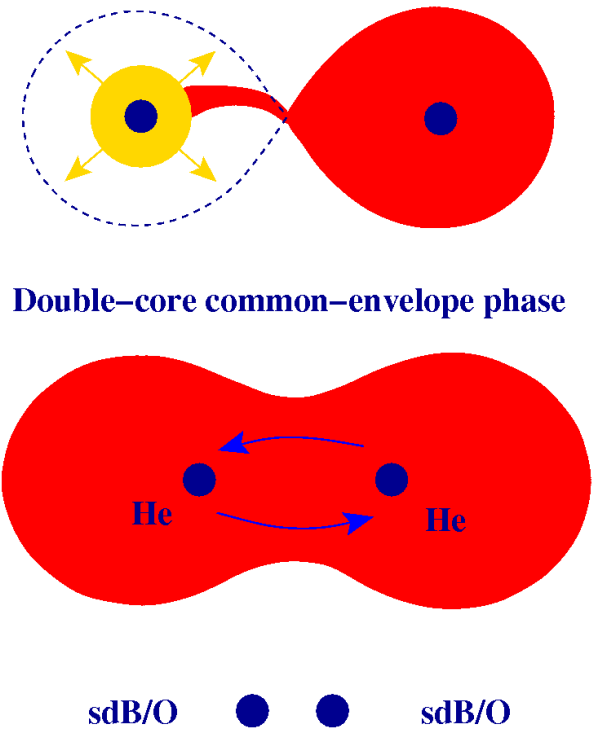, width=8cm}
\caption{ \label{fig:double_core_schematic}
 The double-core common-envelope channel. In a binary with
  binary components of comparable mass, the primary fills its Roche
  lobe after the secondary has completed its main-sequence phase and
  has developed a helium core. This leads to a common-envelope phase,
  where the helium cores of the two stars spiral inside a
  common-envelope formed from their joint envelopes. After the
  ejection of the common envelope, the system may appear as a binary
  consisting of two hot subdwarfs.}
\end{figure}

\section[]{The double He-rich subdwarf stars}
\label{sec:DoubleCore}

The model described by this paper so far can explain the population of
observed single He-rich sdO stars quite well. However, there are
two known \emph{double} He-rich hot subdwarf stars, i.e. binaries
where both components are He-sdO or He-sdB stars. Interestingly, 
both of these systems contain two hot subdwarfs that appear to be
He-rich.

The best published example of such a system is PG 1544+488 (Ahmad et
al.\ 2004), which contains two He-rich hot subdwarf stars. The published
mass ratio for that system is $1.7\pm0.2$, though recent data may
bring this value closer to 1 (Simon Jeffery, priv.\ comm.). The orbital
period is 0.48 d.  Lisker et al.\ (2004) also report the existence
of a system containing two `very similar' He-rich sdO stars (HE
0301-3039; see also Stroeer et al 2007).

A second formation channel is needed to explain these systems: we
suggest that these systems are the products of a 
double-core common-envelope (CE) phase (e.g. Brown 1995; Belczynski \&
Kalogera 2001; Dewi, Podsiadlowski \& Sena 2006).
Double-core CE evolution is a special case of common-envelope
evolution where the envelopes of \emph{both} stars in a binary are
simultaneously ejected as both stellar cores spiral inwards inside an
envelope produced by the union of their envelopes. This produces a
close binary containing the exposed cores of both original stars (see
Fig.~\ref{fig:double_core_schematic}).

The onset of double-core evolution, should it occur, is not driven by
the classical dynamical mass-tranfer instability, since the initial
mass ratio of the two binary components has to be close to one, which
generally leads to dynamically stable mass transfer. However, mass
transfer onto the secondary, which is already trying to expand as a
[sub-]giant, causes it to swell further. Both stars are then trying to
overfill their Roche lobes, leading to the spiral-in and ultimately
the ejection of the shared envelope. Double-core evolution has not
been proven to exist in nature. However, if it is responsible for the
production of double hot subdwarf binaries, then this may have
potentially important implications for the production of double
neutron-star binaries through an analogous channel (e.g.\ Dewi et
al.\ 2006).

The parameter space which is expected to lead to double-core CE
evolution is quite small. The two stars in the binary must have almost
equal initial masses, as the secondary must have already left the main
sequence by the time the primary fills its Roche lobe whilst the
primary should fill its Roche lobe before the tip of the first giant
branch.  In order to form two hot subdwarfs, there is a further
constraint: both stars must be able to ignite helium. The consequence
of this constraint depends on the initial mass of the stars; there are
two regimes which divide approximately into whether the secondary
ignites helium degenerately or non-degenerately.

\begin{figure}
\epsfig{file=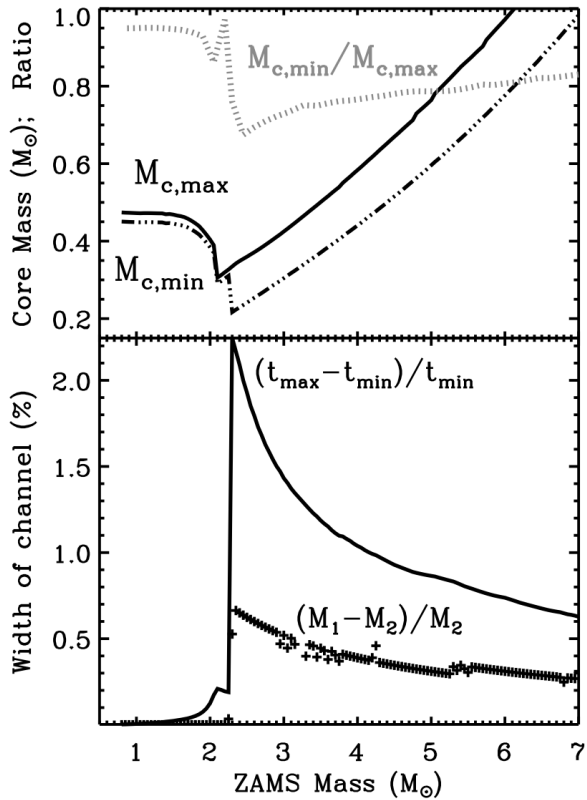, width=8cm}
\caption{Parameter constraints for the double-core channel. The upper
  panel shows the maximum and minimum core masses (respectively the solid and broken
  black curve) that could ignite helium after the
  removal of the stellar envelope in the double-core channel 
  (as a function of zero-age main
  sequence [ZAMS] mass, $M_{\rm ZAMS}$; see also Han et
  al.\ 2002). The broken grey curve shows the ratio
  of those core masses, i.e.\ it gives an estimate for the most extreme hot subdwarf
  mass ratios which could be produced by the double-core channel.
  The lower panel illustrates how the various constraints
  relate to time and mass-ratio constraints. The solid curve shows the
  fractional lifetime difference between $t_{\rm min}$ and $t_{\rm
    max}$, the minimum and maximum age for a given $M_{\rm ZAMS}$ when
  the primary is able to enter into a double-core CE phase.  We do not 
  calculate the energetics of the common-envelope ejection phase. The
  crosses indicate how this lifetime constraint translates into a
  fractional mass difference constraint for the two stars if they are
  \emph{both} to be able to become helium-burning hot subdwarfs
  simultaneously in this channel.  The scatter in the crosses is numerical noise. 
  At best, the initial masses of the
  stars must be within $\approx 0.5$\%. Instances of double-core
  evolution after which both stars can \emph{degenerately} ignite
  helium are highly improbable.
\label{fig:doublecore}
}
\end{figure}

If the secondary is to ignite helium degenerately, its core mass must
be within $\sim$ 5\% of the core mass at the tip of the giant branch
when the envelope is removed (Han et al.\ 2002).  This requirement
also applies at the lower end of the mass range when non-degenerate
ignition is possible.  The likelihood of both components being within
this range at contact is almost negligible, as can be seen from
Fig.~\ref{fig:doublecore}, which quantifies the parameter
constraints for this channel.

If we neglect the unlikely possibility that the secondary ignites
helium degenerately, both components will be igniting helium
non-degenerately.  According to Han et al.\ (2002), stars more massive
than 2.265 $\rm M_{\odot}$ will, at solar metallicity, ignite helium
non-degenerately even if they lose their envelopes in the Hertzsprung
gap.\footnote{At z=0.004, this threshold drops to 2.0 $\rm
  M_{\odot}$; hence adopting a lower metallicity seems likely to increase the birthrate from this channel.}  
This allows a wider range of parameter space to
potentially produce double-hot-subdwarf systems. Figure
\ref{fig:doublecore} shows the width of the non-degenerate double-core
channel. At the lower end of this mass range, the mass of the
secondary has to be within $\approx 0.5\,$\% of the mass of the
primary, declining to $\approx 0.3$\,\%.  This is narrow but not
completely negligible. If we make the standard assumption that the
probability distribution of mass ratios is $p(M_{2}/M_{1}) \propto
M_{2}/M_{1}$ (where $M_{2}/M_{1} \leq 1$), then, for non-degenerate
ignition, around 1\,\% of systems in the first common-envelope channel
from Han et al.\ (2002, 2003) meet this criterion.
However in the preferred population model of Han et al. (2003), only
$\approx$7\% of the close sdB binaries are produced via non-degenerate
ignition in the first CE channel.\footnote{Only 12\% of the sdBs from
  their first common-envelope channel experience non-degenerate
  ignition.} Assuming that half of the known sdBs are in close
binaries, then only one double-core system is born for every $\sim$3000 normal sdBs.  If that estimate is correct, then the birthrate from this channel seems somewhat lower than would comfortably account for seeing two such systems amongst several hundred known sdB stars. However, the two known systems consitute a very small sample size, and we do not attempt to account for observational selection effects.

Figure \ref{fig:doublecore} also shows that hot subdwarf systems
produced from this channel should have mass ratios $1.0 \leq
M_{1}/M_{2} \lessapprox 1.3$, based on the ratio of allowed core
masses.\footnote{Note that the definition of the core mass here is
  based on composition (i.e., $\sim$complete H-exhaustion), which is
  somewhat approximate for intermediate-mass stars at
  the end of the main sequence.}

A third possibility is that the core of the secondary star does not
ignite helium at all. Then the observed system would not contain two
He-burning stars, but one He-burning star and one post-RGB star that
is simply a hot young WD, cooling towards the main WD cooling sequence
(HD 188112 seems to be an example of such a non-He-burning hot
subdwarf; see, e.g., Heber et al.\ 2003; Stroeer et al.\ 2007).  The
lifetime of this cooling phase is short ($\sim 1-10$ Myr, depending on
its mass; Driebe et al.\ 1998; Heber et al. 2003). It seems unlikely to us
that the non-He-burning secondary would resemble the primary for long
enough to reproduce the systems we are considering here, but this
possibility cannot be totally excluded.

\subsection{A non-double-core channel for dual hot subdwarfs}

The double-core CE evolution channel outlined above is not the only
possible way to make a binary containing two hot subdwarfs. Rappaport
et al.\ (2009) have outlined the evolutionary past and
future of the spectroscopic binary Regulus ($\alpha$ Leonis). The
current low-mass ($\sim$ 0.3 $\rm M_{\odot}$) component could
potentially be a very low-mass sdB star.  In one of the possible paths
for the future evolution of the system, the core of Regulus (the
current main-sequence star with mass $\sim$ 3.4 $\rm M_{\odot}$) is
exposed as a second sdB star, of mass $\sim$ 0.5 $\rm M_{\odot}$. In
such a scenario, the lower-mass sdB star could easily be still burning
helium after the second one has been formed.  If the current low-mass
star did not manage to ignite helium, then a system with only
marginally different initial conditions should be able to do so. The
Regulus-like systems would produce a hot subdwarf binary with a mass
ratio far from unity ($\sim 0.5/0.3 \approx1.7$), similar to the
published mass ratio of PG 1544+488. 

There may well be even more channels which can produce double hot
subdwarfs as binary evolution allows for a rich range of
possibilities.  However, the double-core channel tends to produce
systems with mass ratios approaching one with minimal appeal to
fine-tuning.  In addition, one distinguishing feature of both PG
1544+488 and HE 0301-3039 is that they seem to be He-rich. For a
Regulus-like channel it is not so obvious why this should produce
abnormal sdB stars, as they are simply a combination of normal
formation channels. Hence it seems reasonable to ask whether
double-core evolution might somehow tend to produce He-rich subdwarfs.

\begin{figure}
\epsfig{file=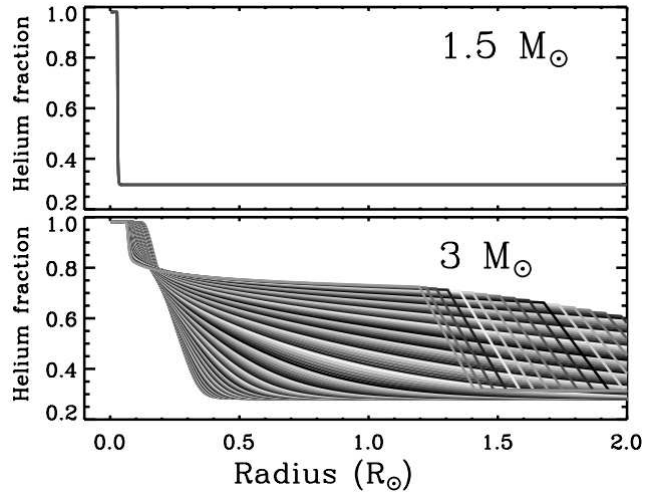, width=8.5cm}
\caption{Comparison of the helium composition profiles of low- and
  intermediate-mass stars (1.5 and 3 $\rm M_{\odot}$, respectively)
  during the evolutionary stage(s) in which they could lose their
  envelopes and still ignite helium in their cores. Many different profiles 
  are shown for the 3 $\rm M_{\odot}$ star, so a greyscale is used to 
  separate adjacent curves. The displayed
  radius range displayed highlights the region near the core which is
  likely to become exposed after the ejection of the envelope during
  double-core evolution. The 1.5 $\rm M_{\odot}$ star has a sharp
  transition between its He-rich core and hydrogen-rich
  envelope. In contrast, the structure of the 3 $\rm M_{\odot}$ star
  allows a relatively high helium abundance outside the core during
  the period when double-core evolution might operate.}
\label{fig:compositionprofile}
\end{figure}

\subsection{Helium rich subdwarfs from double-core evolution?}

As explained above, the double-core channel strongly favours
non-degenerate helium ignition and hence intermediate-mass stars (see
Fig.~\ref{fig:doublecore}). Figure \ref{fig:compositionprofile}
shows that, during the relevant portion of their evolution,
intermediate-mass stars can have extended regions outside their cores
which have high helium abundances, whilst low-mass stars do not. This
seems to provide a natural reason why the known double subdwarf stars
are He-rich.\footnote{An alternative possibility, which we cannot rule
  out at this stage, is that the process of envelope ejection during
  the double-core CE phase is somehow systematically different to that
  of standard common-envelope evolution.}

If intermediate-mass stars tend to produce He-rich hot subdwarfs, then the mass distribution of that subdwarf population should be less strongly peaked than the general sdB population. This is because those sudwarfs from intermediate-mass stars can ignite helium non-degenerately, unlike those low-mass stars which experience the helium flash.

\section[]{Summary and Conclusions}

We have calculated the evolution of post-sdB WD stars after they have
accreted from (or merged with) a helium WD companion.  These stars
burn helium in a shell around the core, but spend a large part of
their evolution with properties mainly determined by the sdB
mass. Their later radius evolution is mainly determined by the
mass of helium gained from their WD companions.  The only major
uncertainty in our modelling is in the treatment of the merger phase
and the degenerate He ignition, but the merger calculations of Saio \&
Jeffery (2000, 2002) seem to support our assumptions.

This demonstrates that one of the major binary channels for
the formation of sdB stars considered by Han et al.\ (2002, 2003)
\emph{predicts} objects which resemble He-rich sdO stars. This subset
of sdO stars `cannot be explained with canonical single star
evolutionary models' (Stroeer et al.\ 2007). An extension of this
model could include mergers of He WDs with other low-mass CO WDs;
however, massive sdB stars which leave remnants resembling normal-mass
CO WDs probably do not produce He-sdOs (see Saio \& Jeffery 2002).  
He-sdOs produced by this scenario have relatively high luminositites
compared to sdB stars, which may help to distinguish between this and
alternative scenarios (e.g. Saio \& Jeffery 2000; Miller Bertolami et
al.\ 2008).

More speculatively, we have also argued that systems containing two
hot subdwarfs (e.g., PG 1544+488, Ahmad et al.\ 2004; HE 0301-3039, Lisker
et al.\ 2004) could form through double-core common-envelope
evolution. 
This generally requires that the subdwarfs have similar masses and
favours intermediate-mass progenitors. Since intermediate-mass giants
have a rather different chemical profile, this may also naturally explain
why hot subdwarfs in these systems are preferentially He-rich.

\section*{Acknowledgements}

We thank the participants at the Bamberg \& Shanghai Hot Subdwarf
conferences for enjoyable discussions, in particular Stephan Geier, 
Uli Heber, Simon Jeffery, Gijs Nelemans \& Roy {\O}stensen. We also
thank Simon Jeffery as the referee; his insightful comments helped improve this
work. SJ has been partially supported by PPARC grant PPA/G/S/2003/00056, 
China National Postdoc Fund Grant No.\ 20090450005 and the 
National Science Foundation of China Grants No.\ 10950110322 and 10903001. 
ZH is supported by the Natural Science Foundation of China under 
Grant No.\ 10821061 and the Chinese Academy of Sciences under 
Grant No.\ KJCX2-YW-T24. Some of this research was done during the KITP
programme `Formation and Evolution of Globular Clusters', and so was
partially supported by the National Science Foundation 
under Grant No. NSF PHY05-51164.

\label{lastpage}

\begin{thebibliography}{}

\bibitem{AJF04} Ahmad A., Jeffery C.S., Fullerton A.W., 2004, A\&A, 418, 275

\bibitem{BK01} Belczynski K., Kalogera V., 2001, ApJ, 550, L183

\bibitem{BSWN07} Bildsten L., Shen K.J., Weinberg N.N., Nelemans G., 2007, ApJ, 662, L95

\bibitem{B95} Brown G.E., 1995, ApJ, 440, 270

\bibitem{Betal01} Brown T.M., Sweigart A.V., Lanz T. Landsman W.B., Hubeny I., 2001, ApJ, 562, 368

\bibitem{DPS06} Dewi J.D.M., Podsiadlowski Ph., Sena A., 2006, MNRAS, 368, 1742

\bibitem{DSBH98} Driebe T., Sch\"{o}nberner D., Bl\"{o}cker T., Herwig F., 1998, A\&A, 339, 123

\bibitem{PPE71} Eggleton P.P., 1971, MNRAS, 151, 351

\bibitem{FST98} Fruchter A.S., Stinebring D.R., Taylor J.H., 1988, Nature, 333, 237

\bibitem{GSL86} Green R.F., Schmidt M., Liebert J., 1986, ApJS, 61, 305

\bibitem{GKH98} Groth H.G., Kudritzki R.P., Heber U., 1985, A\&A, 152, 107

\bibitem{Ha09} Han Zh., 2008, A\&A, 484, L31

\bibitem{HW99} Han Zh., Webbink R.F., 1999, A\&A, 349, L17

\bibitem{HPE94} Han Zh., Podsiadlowski Ph., Eggleton P.P., 1994, MNRAS, 270, 121

\bibitem{HPLG07} Han Zh., Podsiadlowski Ph., Lynas-Gray A.E., 2007, MNRAS, 380, 1098

\bibitem{HPMMI02} Han Zh., Podsiadlowski Ph., Maxted P.F.L., Marsh T.R., Ivanova N., 2002, MNRAS, 336, 449

\bibitem{HPMM03} Han Zh., Podsiadlowski Ph., Maxted P.F.L., Marsh T.R., 2003, MNRAS, 341, 669

\bibitem{He86} Heber U., 1986, A\&A, 155, 33

\bibitem{He08} Heber U., 2008, in Werner K., Rauch T., eds., ASP Conf. Vol. 391, {\it Hydrogen-deficient Stars.} Astron. Soc. Pac.,  San Francisco, p. 245

\bibitem{He09} Heber U., 2009, ARA\&A, 47, 211

\bibitem{HETN03} Heber U., Edelmann H., Lisker T., Napiwotzki R., 2003, A\&A, 411, L477

\bibitem{HHSOHD06} Heber U., Hirsch H., Stro\"{e}r A., O'Toole S., Haas S., Dreizler S., 2006, Baltic Astronomy, 15, 91

\bibitem{HH09} Hirsch H., Heber U., 2009, J.\ Phys.\ Conf.\ Ser., 172, 012015 

\bibitem{I09} Iben I., 1990, ApJ, 353, 215

\bibitem{KW90} Kippenhahn R., Weigert A., 1990, {\it Stellar Structure and Evolution,} Springer-Verlag, Berlin

\bibitem{LBSHL04} Lanz T., Brown T.M., Sweigart A.V., Hubeny I., Landsman W.B., 2004, ApJ, 602, 342

\bibitem{Lisker2004} Lisker T., Heber U., Napiwotzki R., Christlieb N., Reimers D., Homeier D., 2004, Ap\&SS, 291, 351

\bibitem{Lisker2005} Lisker T., Heber U., Napiwotzki R., Christlieb N., Han Zh., Homeier D., Reimers D., 2005, A\&A, 430, 223

\bibitem{MBAUW08} Miller Bertolami M.M., Althaus L.G., Unglaub K., Weiss A., 2008, A\&A, 491, 253

\bibitem{Metal07} Moehler S., Dreizler S., Lanz T., Bono G., Sweigart A.V., Calamida A., et al., 2007, A\&A, 475, L5 

\bibitem{Netal04} Napiwotzki et al., 2004, Ap\&SS, 291, 321

\bibitem{Netal10} Naslim N., Jeffery C.S., Ahmad A., Behara N.,
  \c{S}ahin T., Astrophysics and Space Science Online First, DOI:10.1007/s10509-010-0334-x

\bibitem{NT98} Nelemans G., Tauris T., 1998, A\&A, 335, L85

\bibitem{NS98} Nomoto K., Saio H., 1998, ApJ, 500, 388

\bibitem{Petal95} Pols O.R., Tout C.A., Eggleton P.P., Han Z., 1995, MNRAS, 274, 964

\bibitem{Petal98} Pols O.R., Schr\"{o}der K.-P., Hurley J.R., Tout C.A., Eggleton P.P., 1998, MNRAS, 298, 525

\bibitem{RPH09} Rappaport S., Podsiadlowski Ph., Horev I., 2009, ApJ, 698, 666

\bibitem{SBKL94} Saffer R.A., Bergeron P., Koester D., Liebert J., 1994, ApJ, 432, 351

\bibitem{SJ00} Saio H., Jeffery C.S., 2000, MNRAS, 313, 671

\bibitem{SJ02} Saio H., Jeffery C.S., 2002, MNRAS, 333, 121

\bibitem{SW67} Sandage A., Wildey R., 1967, ApJ, 150, 469

\bibitem{SB09} Shen K.J., Bildsten L., 2009, ApJ, 699, 1365

\bibitem{S98} Soker N., 1998, AJ, 116, 1308

\bibitem{Stroeer07} Stro\"{e}r A., Heber U., Lisker T., Napiwotzki R., Dreizler S., Christlieb N., Reimers D., 2007, A\&A, 462, 269

\bibitem{S97} Sweigart A.V., 1997, The Third Conference on Faint Blue Stars (L. Davis Press) p.3; arxiv:astro-ph/9708164

\bibitem{TP74} Trimble V., Paczynski B., 1974, A\&A, 22, 9

\bibitem{vdB67} van den Bergh S., 1967, AJ, 72, 70

\bibitem{YD097} Yi S., Demarque P., Oemler A., Jr., 1997, ApJ, 486, 201

\bibitem{YDO98} Yi S., Demarque P., Oemler A., Jr., 1998, ApJ, 492, 480

\bibitem{ZCH09} Zhang X., Chen X., Han Z., 2009, A\&A, 504, L13


\end{thebibliography}
\end{document}